\documentclass[12pt]{article}
\usepackage[dvips]{graphicx}
\usepackage{epsfig}
\usepackage{latexsym,amsmath,amsfonts,amssymb}
\usepackage{mathtools} 
\usepackage{slashed} 
\usepackage{blkarray}
\usepackage[all]{xy} 
\usepackage[makeroom]{cancel}
\usepackage[latin1]{inputenc}
\usepackage[american]{babel}
\usepackage[dvips]{graphicx}
\usepackage{bbm}
\usepackage{color}
\usepackage[unicode]{hyperref}
\def\6{\partial}

\newcommand{\be}{\begin{equation}}
\newcommand{\ee}{\end{equation}}
\newcommand{\beq}{\begin{equation}}
\newcommand{\eeq}{\end{equation}}
\newcommand{\bea}{\begin{eqnarray}}
\newcommand{\eea}{\end{eqnarray}}
\newcommand{\nn}{\nonumber \\}

\newcommand{\ba}{\begin{eqnarray}}
\newcommand{\ea}{\end{eqnarray}}

\newcommand{\beqs}{\begin{eqnarray}}
\newcommand{\eeqs}{\end{eqnarray}}
\newcommand{\bal}{\begin{aligned}}
\newcommand{\eal}{\end{aligned}}

\def\lbldef#1#2{\expandafter\gdef\csname #1\endcsname {#2}}

\def\href#1#2{#2}

\newcommand{\ber}{\begin{eqnarray}}
\newcommand{\eer}{\end{eqnarray}}

\newcommand{\beqar}{\begin{eqnarray}}

\newcommand{\eeqar}{\end{eqnarray}}

\newcommand{\dsl}
   {\kern.06em\hbox{\raise.15ex\hbox{$/$}\kern-.56em\hbox{$\partial$}}}

\newcommand{\eeqarr}{\end{eqnarray}}
\newcommand{\ZZ}{{\rm \kern 0.275em Z \kern -0.92em Z}\;}

\def\CC{{\mathchoice
{\rm C\mkern-8mu\vrule height1.45ex depth-.05ex
width.05em\mkern9mu\kern-.05em}
{\rm C\mkern-8mu\vrule height1.45ex depth-.05ex
width.05em\mkern9mu\kern-.05em}
{\rm C\mkern-8mu\vrule height1ex depth-.07ex
width.035em\mkern9mu\kern-.035em}
{\rm C\mkern-8mu\vrule height.65ex depth-.1ex
width.025em\mkern8mu\kern-.025em}}}

\def\RR{{\rm I\kern-1.6pt {\rm R}}}

\def\ZZ{{\rm Z}\kern-3.8pt {\rm Z} \kern2pt}
\def\IB{\relax{\rm I\kern-.18em B}}
\def\ID{\relax{\rm I\kern-.18em D}}
\def\II{\relax{\rm I\kern-.18em I}}
\def\IP{\relax{\rm I\kern-.18em P}}

\newcommand{\bear}{\begin{eqnarray}}
\newcommand{\eear}{\end{eqnarray}}

\def\6{\partial}

\newfont{\namefont}{cmr10}
\newfont{\addfont}{cmti7 scaled 1440}
\newfont{\boldmathfont}{cmbx10}
\newfont{\headfontb}{cmbx10 scaled 1728}

\newcommand{\dd}{\textrm{d}}

\allowdisplaybreaks[1]

\numberwithin{equation}{section}

\begin{document}
\begin{titlepage}

\vfill
\begin{flushright}
APCTP Pre2018-001 
\end{flushright}

\vfill

\begin{center}
   \baselineskip=16pt
   {\Large \bf On non-Abelian T-duality for non-semisimple groups}
   \vskip 2cm
     Moonju Hong$^a$, Yoonsoo Kim$^a$, Eoin \'O Colg\'ain$^b$
       \vskip .6cm
             \begin{small}
               \textit{$^a$ Department of Physics, Postech, Pohang 37673, Korea}
               
               \vspace{3mm} 
               
               \textit{$^b$ Asia Pacific Center for Theoretical Physics, Postech, Pohang 37673, Korea}
  
             \end{small}
\end{center}

\vfill \begin{center} \textbf{Abstract}\end{center} \begin{quote}
We revisit non-Abelian T-duality for non-semisimple groups, where it is well-known that a mixed gravitational-gauge anomaly leads to $\sigma$-models that are scale, but not Weyl-invariant. Taking into account the variation of a non-local anomalous term in the T-dual $\sigma$-model of Elitzer, Giveon, Rabinovici, Schwimmer \& Veneziano, we show that the equations of motion of generalized supergravity follow from the $\sigma$-model once the Killing vector $I$ is identified with the trace of the structure constants. As a result, non-Abelian T-duals with respect to non-semisimple groups are solutions to generalized supergravity. We illustrate our findings with Bianchi spacetimes. 
\end{quote} \vfill

\end{titlepage}
\setcounter{equation}{0}


\section{Introduction}
Following Buscher's seminal work on T-duality \cite{Buscher:1987sk}, a generalisation to non-Abelian isometries was quickly proposed \cite{delaOssa:1992vci} \footnote{See \cite{Fridling:1983ha} for earlier examples.}. One striking feature of non-Abelian T-duality is that it breaks isometries, but it also turned out to be novel in other ways. In particular, it was demonstrated that non-Abelian T-duality was not a symmetry of conformal field theory, but rather a symmetry between different theories \cite{Giveon:1993ai}. Moreover, it was noted by Gasperini, Ricci \& Veneziano that the procedure failed to provide a valid supergravity solution for Bianchi V  \cite{Gasperini:1993nz} and Bianchi III \cite{Gasperini:1994du} cosmological models. It was subsequently realised that structure constants with a non-vanishing trace were related to a mixed gravitational-gauge anomaly \cite{Alvarez:1994np, Elitzur:1994ri}, which explained why non-Abelian T-duality was no longer a symmetry of supergravity. Eventually, non-Abelian T-duality was extended to the RR sector \cite{Sfetsos:2010uq} and became a powerful solution generating technique \cite{solgen}, especially for AdS/CFT geometries, where implications for the dual CFTs were explored \cite{AdSCFT}. However, the fate of the non-Abelian T-dual geometry of Bianchi V and III remained a puzzle. 

In recent years, we have witnessed further interest in non-Abelian T-duality, driven by swift developments in integrable $\sigma$-models \cite{Klimcik:2002zj}. We recall that non-Abelian T-duals arise as limits of $\lambda$-deformations of $AdS_p \times S^p$ geometries \cite{Sfetsos:2014cea}, while homogeneous Yang-Baxter deformations \cite{Kawaguchi:2014qwa} can be understood as non-Abelian T-duality transformations \cite{Hoare:2016wsk, Borsato:2016pas} \footnote{Yang-Baxter deformations can be understood as open-closed string maps \cite{Seiberg:1999vs} where the $r$-matrix is the noncommutativity parameter $\Theta$, $\Theta = r$  \cite{Araujo:2017jkb, Araujo:2017enj} (see \cite{vanTongeren:2016eeb} for an earlier observation in a restricted setting).  }. One important by-product of Yang-Baxter deformations was the discovery that there are integrable deformations that are not solutions of usual supergravity, but a modification, called \textit{generalized supergravity} \cite{Arutyunov:2015mqj} (see also \cite{Wulff:2016tju}). The modification is specified by a Killing vector $I$ and bona fide supergravity solutions correspond to $I = 0$. 

With the advent of generalized supergravity \cite{Arutyunov:2015mqj} and the knowledge that homogeneous Yang-Baxter deformations are non-Abelian T-duality transformations \cite{Borsato:2016pas}, it would be surprising if non-Abelian T-duals for non-semisimple groups did not also solve the equations of motion (EOMs) of generalized supergravity (as originally anticipated in \cite{Hoare:2016wsk}). For the Bianchi V geometry this was confirmed recently \cite{Fernandez-Melgarejo:2017oyu}. Here, we extend this observation to Bianchi VI$_h$, a one-parameter family of groups that include Bianchi III $(h=0)$ and Bianchi V $(h=1)$ as special cases. Since Bianchi IV and VII$_h$ give rise to singular supergravity solutions \cite{Batakis:1995kn}, this exhausts all Bianchi cosmologies based on non-semisimple groups. 

For non-Abelian T-duals of Bianchi cosmologies, we observe that the Killing vector $I$ of generalized supergravity is simply the trace of the structure constants. While this agreement may be coincidental, to better understand this feature we return to the T-dual $\sigma$-model of Elitzer, Giveon, Rabinovici, Schwimmer \& Veneziano (EGRSV), which includes the contribution from a non-local anomalous term \cite{Elitzur:1994ri}. A key observation of EGRSV was that non-vanishing $\beta$-functions for Bianchi V could be cancelled by the variation of an additional non-local term with respect to the conformal factor, which they demonstrated explicitly for Bianchi V. Here, we confirm this result for Bianchi III, before providing proof that for any background the one-loop $\beta$-functions of the EGRSV $\sigma$-model agree with the equations of motion of generalized supergravity once the Killing vector is identified with the trace of the structure constants, $I^{i} = f^{j}_{j i}$. 

To get a better grasp of this claim, let us recall that the EGRSV $\sigma$-model is scale invariant, but not Weyl-invariant. Being scale invariant, it is known that the one-loop $\beta$-functions must take the form \cite{Friedan:1980jf}
\bea
\beta_{G_{\mu \nu}} &=& R_{\mu \nu} - \frac{1}{4} H_{\mu \rho \sigma} H_{\nu}^{~ \rho \sigma} + \nabla_{\mu } X_{\nu} + \nabla_{\nu} X_{\mu} , \\
\beta_{B_{\mu \nu}} &=& - \frac{1}{2} \nabla^{\rho} H_{\rho \mu \nu} +X^{\rho} H_{\rho \mu \nu} + \nabla_{\mu} Y_{\nu} - \nabla_{\nu} Y_{\mu}, 
\eea
for arbitrary vectors, $X$ and $Y$. We recall that for $X_{\mu} = \partial_{\mu} \Phi$, $Y_{\mu} =0$, we recover the usual one-loop $\beta$-functions of supergravity \cite{Callan:1985ia}, where it should be noted that the contribution due to the dilaton is a \textit{classical} contribution at the same order as the one-loop \textit{quantum} contributions of the $G_{\mu \nu}$ and $B_{\mu \nu}$ couplings. This happens because the term $R^{(2)} \Phi$ is scale non-invariant at the classical level, while the other couplings lose scale invariance at one-loop. At this point, we could adopt the strategy employed in \cite{Arutyunov:2015mqj} and use an explicit solution to fix the vectors. However, provided one includes the anomaly term in the EGRSV $\sigma$-model, which appears at the same order as the dilaton, and simply varies it with respect to the conformal factor, we shall see that the equations of motion of generalized supergravity can be derived.  

The structure of this short note is as follows. In section \ref{sec:review} we review non-Abelian T-duality with respect to both semisimple and non-semisimple groups.  In section \ref{sec:Bianchi}, we introduce Bianchi cosmologies and describe non-Abelian T-dualities of both Bianchi I and Bianchi II spacetimes, noting in the former case that the matrix inversion inherent to non-Abelian T-duality is simply three commuting Abelian T-dualities. In section \ref{sec:trace}, we demonstrate that non-Abelian T-duals of Bianchi VI$_h$ cosmologies lead to generalized supergravity solutions once the Killing vector is identified with the trace of the structure constants. In the special case where $h=-1$, the trace of the structure constants vanishes and we find a solution to usual supergravity. In section \ref{sec:EGRSV}, we explain why $I^{i} = f^{j}_{j i}$. The EOMs of generalized supergravity can be found in the appendix. 

\section{Review of non-Abelian T-duality} 
\label{sec:review}
In this section we quickly review non-Abelian T-duality. We will do this in two stages: first, we introduce the transformation for semisimple groups without isotropy, before explaining how the T-dual $\sigma$-model is modified for non-semisimple groups. Following \cite{Gasperini:1993nz}, we will tailor the discussion to Bianchi cosmologies from the outset. Consider the 2D string $\sigma$-model 
\be
S = \frac{1}{2 \pi} \int \dd^2 z \,  \left[  \partial X^{\mu} ( G_{\mu \nu} + B_{\mu \nu} ) \bar{\partial} X^{\nu}  + 2 \Phi \partial \bar{\partial} \sigma \right], 
\ee
where $X^{\mu}$, $\mu = 0, \dots, d$, denote the target spacetime coordinates, the couplings $G_{\mu \nu}$, $B_{\mu \nu}$ are symmetric and anti-symmetric, respectively, and correspond to the target space metric and NSNS two-form. $\sigma$ is the worldsheet conformal factor,
\be  
\partial \bar{\partial} \sigma = \frac{1}{4} \sqrt{h} R^{(2)}, 
\ee
where $h$ is the worldsheet metric and $R^{(2)}$ the worldsheet curvature with $\Phi$ denoting the scalar dilaton.

Let us now assume that the target space has an isometry group, where the generators of the corresponding Lie algebra can be expressed in terms of the Killing vectors $K_{i}$ of the target space geometry, 
\be
K_{i} = K^{m}_{i} \partial_{m}, \quad i = 1, \dots, d,  
\ee
where $d$, in addition to being the dimension of the space, is also the dimension of the Lie algebra. The Lie algebra is fully specified by the structure constants, 
\be
[ K_{j}, K_{k} ] = f_{j k}^{i} K_{i}.  
\ee
Dual to the Killing vectors $K_{i}$, one can define Maurer-Cartan one-forms $\sigma_{i}$, which satisfy a related differential condition: 
\be
\label{MC}
\dd \sigma_i = \frac{1}{2} f^i_{jk} \sigma_j \wedge \sigma_k. 
\ee

For Lie algebras of dimension three, $d=3$, we have a complete classification due to Bianchi \cite{Bianchi} and it is this setting in which we will consider non-Abelian T-duality. In fact for each family of symmetries, one can define a corresponding ``Bianchi cosmology", which is a 4D spacetime, parametrised by coordinates $(t, \vec{x})$, where the internal 3D space exhibits the symmetries of the corresponding Bianchi class and whatever warp factors appear only depend on the time direction. Within this class of geometries, the usual Friedmann-Robertson-Walker (FRW) solutions correspond to Bianchi I, V and IX. 

Following \cite{Gasperini:1993nz}, we can factorise the spacetime metric and NSNS two-form so that all spatial dependence $x^i$ is contained in a dreibein $e^{i}_{m}$, 
\be
G_{m n} ( t , \vec{x}) = e^{i}_{m} (\vec{x}) \gamma_{i j}(t) e^{j}_{n} (\vec{x}), \quad B_{m n} (t, \vec{x}) = e^{i}_{m} (\vec{x}) \beta_{ij}(t) e^{j}_{n} (\vec{x}), \quad \gamma_{ij} = \gamma_{ji}, \quad \beta_{ij} = - \beta_{ji}, 
\ee
where we have isolated the spatial directions $m = 1, 2, 3$. We are assuming that $B_{0 m } = G_{0 m } = 0$. At this stage, one performs non-Abelian T-duality by gauging the isometries \cite{Rocek:1991ps}
\be
\partial X^m \rightarrow \partial X^m + A^{l} K^{m}_l, \quad \bar{\partial} X^m \rightarrow \bar{\partial} X^m + \bar{A}^{l} K^{m}_l,  
\ee
in the process introducing a set of pure gauge potentials $A^{l}, \bar{A}^l$. The new action $S'$ is the original action plus an additional contribution, 
\bea
S' = S &+& \frac{1}{2 \pi} \int \dd^2 z \biggl( A^{l} K^{m}_{l} e^{i}_{m} ( \gamma_{ij} + \beta_{ij} ) e^{j}_n \bar{\partial} X^n  + \partial X^m e^{i}_{m} ( \gamma_{ij} + \beta_{ij} ) e^{j}_n \bar{A}^l K^n_l \nn
&+& A^{l} K^{m}_{l} e^{i}_{m} ( \gamma_{ij} + \beta_{ij} ) e^{j}_n \bar{A}^k K^n_k + \tilde{X}_i F^i \biggr), 
\eea
where $F^i$ is the field strength corresponding to $A^i, \bar{A}^i$, $F^i  = \partial \bar{A}^i - \bar{\partial} A^i +  f^i_{jk} A^j \bar{A}^k$  and we have added the Lagrange multiplier to enforce the condition $F^i = 0$. Strictly speaking, integration on $\tilde{X}$ only ensures that $A, \bar{A}$ are pure gauge in spherical worldsheets, in which case, one recovers the original $\sigma$-model. Instead of integrating out the Lagrange multiplier, one can integrate out the gauge potentials to get a dual $\sigma$-model, before subsequently fixing the residual gauge symmetry. A convenient gauge choice is simply to take $X^m$ to be constant, which yields the dual action 
\be
\label{dual_action}
\tilde{S} = \frac{1}{2 \pi} \int \dd^2 z \left[  \partial X^{0} G_{00}  \bar{\partial} X^{0} + \partial \tilde{X}_{i} M^{i j } \bar{\partial} \tilde{X}_j  + (2  \Phi + \ln \det M) \partial \bar{\partial} \sigma \right], 
\ee
where we have defined the matrix 
\be
\label{map}
M = (\gamma + \beta + \kappa)^{-1}
\ee
in terms of the anti-symmetric matrix $\kappa$:
\be
\label{kappa}
\kappa_{ij} \equiv f^{k}_{ij} \tilde{X}_k. 
\ee
As is evident from the dual action (\ref{dual_action}), the T-dual metric $\tilde{G}$ and NSNS two-form $\tilde{B}$ are read off from the symmetric and anti-symmetric components of $M$, $\tilde{G} + \tilde{B} = M$ and it is worth noting that the Lagrange multipliers become the T-dual coordinates. Furthermore, the dilaton shift
\be
\label{dilaton_shift}
\Phi \rightarrow \Phi + \frac{1}{2} \ln \det M, 
\ee 
is the result of a Jacobian factor from integrating out the gauge fields \cite{Buscher:1987sk}. When $\kappa = 0$, so that all the structure constants vanish, this transformation reduces to the usual Buscher rules for Abelian T-duality. It is clear that one can generate complicated geometries through this map, so for simplicity we focus on examples where $\beta = 0$. 

The above treatment is adequate for semisimple groups. For non-semisimple groups, there is a non-local anomalous contribution to the $\sigma$-model \cite{Elitzur:1994ri}, 
\be
S_{\textrm{non-local}} =  - \frac{1}{8 \pi}  f_{j i}^{j} \int \dd^2 z \left( \frac{1}{\partial} A^i  + \frac{1}{\bar{\partial}} \bar{A}^i \right) \sqrt{h} R^{(2)}
\ee
which once the gauge fields are integrated out leads to a new dual action, 
\be
\label{dual_action2}
\tilde{S} = \frac{1}{2 \pi} \int \dd^2 z \left[  \partial X^{0} G_{00}  \bar{\partial} X^{0} + ( \partial \tilde{X}_{i} - f^{k}_{ki} \partial \sigma ) M^{i j } (\bar{\partial} \tilde{X}_j  + f^{l}_{lj} \bar{\partial} \sigma) + (  2 \Phi +  \ln \det M) \partial \bar{\partial} \sigma \right]. 
\ee
It should be noted that irrespective of the group, the transformation of the metric, NSNS two-form and dilaton follow the same prescription, but there is a difference in the dual $\sigma$-model, which will be important later. 

\section{Bianchi Cosmology}
\label{sec:Bianchi} 
Having explained the fundamental map (\ref{map}) at the heart of Buscher procedure, we put it to work on Bianchi cosmologies. Our main focus will be discussing non-Abelian T-duals of Bianchi cosmologies where the trace of the structure constants is non-vanishing. Before discussing these more exotic examples, we will begin by introducing the Bianchi spacetimes and discussing both Abelian and non-Abelian T-duality in the simpler setting of Bianchi I and Bianchi II.  

We will follow the description of the spacetimes presented in \cite{Batakis:1995kn}, where a general family of supergravity solutions, namely spacetimes supported both by the scalar dilaton $\Phi$ and NSNS two-form $B$, were presented. As stated earlier, since the NSNS two-form only complicates the non-Abelian T-duality, we will further restrict the solutions presented in \cite{Batakis:1995kn} to $B=0$. With this restriction, the solutions are expected to agree with \cite{Gasperini:1994du}. Before beginning, we warn the reader that our dilaton is not the dilaton $\phi$ presented in \cite{Batakis:1995kn}, but instead $\Phi = - \frac{1}{2} \phi$. 

To begin, let us consider the Bianchi spacetime \cite{Batakis:1995kn}, 
\be
\label{Bianchi_metric}
\dd s^2 = - a_1^2 a_2^2 a_3^2 e^{-4 \Phi} \, \dd t^2 + a_1^2 \, \sigma_1^2 + a_2^2 \, \sigma_2^2 + a_3^2 \, \sigma_3^2, 
\ee
where $a_i$ and $\Phi$ are functions of $t$. Note, it is more usual to fix the gauge so that $g_{tt} = - 1$. In contrast, the above form is unorthodox, but the advantage of the rescaled temporal direction is that the dilaton equation is simplified. Before discussing it, let us note that when $I = 0$ and $B=0$, the EOM for the NSNS two-form (\ref{B_EOM}) is trivially satisfied, so we only need to discuss the Einstein equation (\ref{Einstein}) and the dilaton EOM (\ref{dilaton_EOM}). 

Setting $I = B = 0$ in the Einstein equation (\ref{Einstein}) and contracting, we get, 
\be
R + 2 \nabla^2 \Phi = 0. 
\ee
Combining with the dilaton equation (\ref{dilaton_EOM}), we can eliminate the Ricci scalar $R$ to get 
\be
\nabla_{\mu} \nabla^{\mu} \Phi - 2 \partial_{\mu}  \Phi \partial^{\mu} \Phi =  \frac{1}{\sqrt{-g}} \partial_{\mu} ( e^{-2 \Phi} \sqrt{-g} g^{\mu \nu} \partial_{\nu} \Phi )   = 0. 
\ee

Since $\Phi$ is assumed to be only a function of $t$, given the above spacetime (\ref{Bianchi_metric}), we arrive at an easily solved equation: 
\be
\label{linear_dilaton}
\partial_t^2 \Phi = 0, \quad \Rightarrow \quad \Phi = \beta t, 
\ee
where we have exploited the freedom to shift $\Phi$ to remove a constant. We now turn to the Einstein equation. To solve the Einstein equation, we need to consider a given Bianchi class with specific Maurer-Cartan one-forms. For Bianchi I spacetimes, the Maurer-Cartan forms are simply $\sigma_1 = \dd x, \sigma_2  = \dd y, \sigma_3 = \dd z$, which are all closed and therefore from (\ref{MC}) all the structure constants vanish. In this case, the functions $a_i$ are  \cite{Batakis:1995kn}
\be
a_i = e^{ p_i t}, 
\ee
where $p_i$ are constants and we have absorbed additional constants by redefining the coordinates $x, y$ and $z$. The final equation is the Einstein equation in the time direction, which holds once the constants we have introduced satisfy the following equation \footnote{There appears to be a mistake in \cite{Batakis:1995kn}. The three-form $H = \dd B = A \sigma_1 \wedge \sigma_2 \wedge \sigma_3$, so $A= 0$ should recover our result, but instead the quoted result is $\sum_{i < j} p_i p_j = 0$. Since this implies the dilaton does not back-react, one concludes there is a typo.}
\be
\label{E00_eq}
E_{tt} = \sum_{i < j} p_i p_j - 2 \beta \sum_{i=1}^3 p_i + 2 \beta^2 = 0. 
\ee
Let us quickly summarise the Bianchi I spacetime. The final solution is 
\bea
\label{BianchiI_sol}
\dd s^2 &=& -  e^{2 (p_1 + p_2 + p_3 - 2 \beta)t } \, \dd t^2 + e^{2 p_1 t} \, \sigma_1^2 +  e^{2 p_2 t} \, \sigma_2^2 + e^{2 p_3 t} \, \sigma_3^2, \nn
\Phi &=& \beta t, 
\eea
where the constants are subject to the single condition (\ref{E00_eq}). When $\beta = 0$, rescaling the time direction, one can confirm that we recover the Kasner solution \cite{Kasner:1921zz}, as expected. 

\subsection{Bianchi I: Abelian as a non-Abelian T-duality}
The solution (\ref{BianchiI_sol}) is the simplest Bianchi I solution supported only by a scalar field. The solution has three Killing vectors, $\partial_{x}, \partial_y$ and $\partial_z$, which generate translations in the spatial directions. The Killing vectors commute with one another so all structure constants associated with the Lie algebra vanish. 

We will now perform three commuting Abelian T-dualities along the isometry directions, but we will do so from the perspective of a non-Abelian T-duality transformation with vanishing structure constants. We will see that this simply corresponds to inverting a matrix. As discussed in section \ref{sec:review}, the matrix associated to the Maurer-Cartan one-forms is 
\be
\gamma = \left( \begin{array}{ccc} e^{2 p_1} & 0 & 0 \\ 0 & e^{2 p_2} & 0 \\  0 & 0 &  e^{2 p_3}\end{array} \right). 
\ee
We observe that it is symmetric, so the inverse is also symmetric. As a result, we will not generate a $B$-field and the components of the inverse matrix correspond to the T-dual metric: 
\be
\label{I_metric_dual}
\dd s^2_{\textrm{dual}} = -  e^{2 (p_1 + p_2 + p_3 - 2 \beta) t} \, \dd t^2 + e^{- 2 p_1 t} \, \dd x^2 +  e^{-2 p_2 t} \, \dd y^2 + e^{-2 p_3 t} \, \dd z^2, 
\ee
where we have introduced $(\tilde{X}_1, \tilde{X}_2, \tilde{X}_3) = (x, y, z)$ as the dual coordinates, in line with usual practice. However, it should be noted that we have the freedom to label the dual coordinates as we please and we are still guaranteed to produce a solution. Recall that we have gauge fixed so the original coordinates disappeared in the transformation leaving the Lagrange multipliers to take their place. 

To read off the transformation for the dilaton, we can either use the shifted expression (\ref{dilaton_shift}), or simply note that the density $e^{-2 \Phi} \sqrt{-g} $ is invariant. Using the latter, while neglecting the $g_{tt}$ term as it does not change, we get 
\be
e^{(p_1 + p_2 + p_3 -  2 \beta ) t}  = e^{-2 \Phi} e^{-(p_1 + p_2 + p_3) t}, 
\ee
which implies the T-dual dilaton is 
\be
\label{I_dilaton_dual}
\Phi = (\beta - p_1 - p_2 - p_3)t. 
\ee
Substituting the new metric (\ref{I_metric_dual}) and dilaton (\ref{I_dilaton_dual}) into the supergravity EOMs, we find that the EOMs are satisfied provided (\ref{E00_eq}) holds. This confirms that we have generated a new solution from old. 

While we have not performed a genuine non-Abelian T-duality, through this example we have recast Abelian T-duality, or more accurately three commuting T-dualities, as a non-Abelian duality transformation where all the structure constants vanish. As we have seen, the Buscher procedure for this simple example reduces to inverting a matrix. This same operation will be at the heart of the subsequent examples, but anti-symmetric components, essentially introduced via $\kappa$ (\ref{kappa}) will generate a $B$-field. 

\subsection{Bianchi II: semisimple warm-up} 
For Bianchi II spacetimes, we consider the same metric (\ref{Bianchi_metric}), but now the one-forms are 
\be
\sigma_1 = \dd x -z \dd y, \quad \sigma_2 = \dd y , \quad \sigma_3 = \dd z.
\ee
The structure constants are obtained from the differential conditions on the one-forms, 
\be
\dd \sigma_1 = \sigma_2 \wedge \sigma_3 \quad \Rightarrow \quad f_{23}^1 = 1. 
\ee
Since the structure constant is traceless, the T-dual geometry will be a solution to usual supergravity. We recall that the dilaton is the same as Bianchi I, $\Phi = \beta t$, and from \cite{Batakis:1995kn} we find that the functions $a_i$ in general can be written as
\bea
a_1 &=& e^{\Phi} \Big( \frac{p_1}{\cosh (p_2 t)} \Big)^{1/2}, \nn
a_2 &=& e^{\Phi} \cosh (p_1 t) ^{1/2} e^{\frac{1}{2}p_2 t} , \nn
a_3 &=& e^{\Phi} \cosh (p_1 t) ^{1/2} e^{\frac{1}{2}p_3 t},
\eea
where some unnecessary constants have been absorbed into coordinates. One can check that EOMs require the condition
\be
\label{Bianchi_II}
p_2 p_3 - p_1^2 = 4\beta ^2. 
\ee

It turns out that performing T-duality corresponds to inverting the matrix,
\be
\gamma + \kappa  = \left( \begin{array}{ccc} 
a_1^2 & 0  & 0 \\ 
0  & a_2^2 & x \\
0 & -x & a_3^2 
\end{array} \right),  
\ee
with $a_i$ as presented above. In contrast to the previous example, our metric now has an anti-symmetric component. We have once again chosen to label the Lagrange multipliers $(\tilde{X}_1, \tilde{X}_2, \tilde{X}_3) = (x, y, z)$, so from $\kappa$ (\ref{kappa}) the 23-component of the matrix is $x$. 

From the inverse matrix, we read off the dual metric and $B$-field, 
\bea
\dd s^2 &=&  - a_1^2 a_2^2 a_3^2 e^{-4 \beta t } \, \dd t^2 + \frac{1}{\Delta} \left( (a_2^2 a_3^2 + x^2) \dd x^2 + a_1^2 a_3^2 \dd y^2 + a_1^2 a_2^2 \dd z^2 \right), \nn
B &=& -\frac{a_1^2 }{\Delta} x \dd y \wedge \dd z,
\eea 
where we have defined 
\be
\Delta = a_1^2 (a_2^2 a_3^2 + x^2). 
\ee
Having started with $\Phi = \beta t$, the transformed dilaton is
\be
\Phi =  \beta t -\frac{1}{2}\log \Delta.
\ee
Substituting the explicit values for $a_i$, one finds that EOMs are satisfied through (\ref{Bianchi_II}). 

\section{Solutions to Generalized Supergravity} 
\label{sec:trace}
As outlined in the introduction, our main motivation is to show that there are Bianchi cosmologies outside of the (Ricci-flat) Bianchi V class  \cite{Fernandez-Melgarejo:2017oyu} that give rise to generalized supergravity solutions under non-Abelian T-duality transformations. We focus on Bianchi VI$_h$ cosmologies as these are the only non-singular solutions where the structure constants have a non-vanishing trace. Interestingly, VI$_h$ is a one-parameter family of groups, which covers both Bianchi III and V, but also includes one group ($h = -1$) where the trace of the structure constants vanishes. Therefore, for Bianchi VI$_h$ $h\neq -1$ we expect to find non-Abelian T-duals that are solutions to generalized supergravity, but for VI$_{-1}$ we anticipate that the T-dual will be a genuine supergravity solution.  We begin with the generic case. 

\subsection{Bianchi VI$_{h}$}
We can import the solution directly from \cite{Batakis:1995kn}. The spacetime takes the same form (\ref{Bianchi_metric}), but we redefine the Maurer-Cartan one-forms as
\be
\sigma_1 = \dd x, \quad \sigma_2 = e^{ h x} \dd y, \quad \sigma_3 = e^{x} \dd z, 
\ee 
and the dilaton is unchanged (\ref{linear_dilaton}). The functions appearing in the metric may be expressed as \cite{Batakis:1995kn}: 
\bea
a_1 &=& e^{\Phi} \left( \frac{p_1}{h+1} \right)^{\frac{(h^2+1)}{(h+1)^2}} \sinh ( p_1 t )^{- \frac{(h^2+1)}{(h+1)^2}} e^{ \frac{(h-1)}{2(h+1)} p_2 t }, \nn
a_2 &=& e^{\Phi}  \left( \frac{p_1}{h+1} \right)^{\frac{h}{(h+1)}} \sinh ( p_1 t )^{- \frac{h}{(h+1)}} e^{ \frac{1}{2} p_2 t }, \nn
a_3 &=& e^{\Phi} \left( \frac{p_1}{h+1} \right)^{\frac{1}{(h+1)}} \sinh ( p_1 t )^{- \frac{1}{(h+1)}} e^{- \frac{1}{2} p_2 t }, \nn
\eea
where $p_i$ denote constants and we have absorbed redundant constants. From these expressions, it is clear that the $h=-1$ case has to be treated separately, which we do in the next section. It can be confirmed that all equations are satisfied once the following condition is satisfied:
\be
\label{VI_cond}
\frac{4(h^2 + h+1)}{(h+1)^2} p_1^2 = p_2^2 + 4 \beta^2. 
\ee
The Maurer-Cartan one-forms satisfy the differential conditions: 
\be
\dd \sigma_1 = 0, \quad \dd \sigma_2 = h \sigma_1 \wedge \sigma_2, \quad \dd \sigma_3 = \sigma_1 \wedge \sigma_3, 
\ee
and the corresponding Killing vectors are respectively, 
\be
K_1 = \partial_x- z \partial_z - h y \partial_y, \quad K_2 = \partial_{y}, \quad K_3 = \partial_{z}. 
\ee
The Killing vectors satisfy the Lie algebra
\be
[K_1, K_2] = h K_2, \quad [K_1, K_3] = K_3.   
\ee
From either the differential forms or the Killing vectors, one can easily identify the structure constants
\be
f_{12}^2 = h, \quad f_{13}^3 = 1. 
\ee
Immediately, one notes that the trace is zero when $h = -1$. With the structure constants at hand, we are set to perform the non-Abelian T-duality. As prescribed earlier, all we have to do is invert the matrix, 
\be
\gamma + \kappa = \left( \begin{array}{ccc} a_1^2 & h y  & z \\ 
-h y  & a_2^2 & 0 \\
-z & 0 & a_3^2 
\end{array} \right) .
\ee
and extract the symmetric and anti-symmetric components: 
\bea
\dd s^2 &=& - a_1^2 a_2^2 a_3^2 e^{-4 \beta t} \dd t^2 +  \frac{1}{\Delta} \biggl( a_2^2 a_3^2 \dd x^2 + (z^2 + a_1^2 a_3^2) \dd y^2 - 2 h y z \dd y \dd z + (h^2 y^2 + a_1^2 a_2^2) \dd z^2 \biggr) \nn
B &=& -\frac{1}{\Delta} \dd x \wedge ( h y a_3^2 \dd y + z a_2^2 \dd z), 
\eea
where we have defined 
\be
\Delta = h^2 y^2 a_3^2 + a_2^2 ( z^2 + a_1^2 a_3^2).
\ee
The change in the dilaton is once again easily read off, giving us 
\be
\Phi = \beta t - \frac{1}{2} \log \Delta. 
\ee
As with the original Bianchi V T-dual \cite{Gasperini:1993nz}, or the later Bianchi III T-dual \cite{Gasperini:1994du}, the metric, NSNS two-form and dilaton do not  satisfy the usual supergravity EOMs on their own. This was the original puzzle posed a quarter century ago. However, once complemented with the appropriate Killing vector, in this case 
\be
I = - ( h + 1) \partial_{x}, 
\ee
it is a straightforward exercise to check that the generalized supergravity EOMs are satisfied. Recalling that VI$_h$ includes Bianchi III and V as special cases, our analysis reduces to the earlier result of \cite{Fernandez-Melgarejo:2017oyu} when $h = 1$. It is worth noting that $I$ is simply the trace of the structure constants, $I^{1} = f^{i}_{i1}$, where we have used the fact that $\tilde{X}_1 = x$.

\subsection{Bianchi VI$_{-1}$}
Once again, we can reproduce the solution from \cite{Batakis:1995kn}. Up to constants, which can be absorbed, the functions may be expressed as 
\bea
a_1 = \sqrt{p_1} \exp \left[ \frac{1}{2} e ^{2 p_2 t} + \left(\frac{p_1}{2} + \beta \right) t \right], \quad a_3 = a_2 = \sqrt{p_2} e^{( \frac{p_2}{2} + \beta) t}, \quad \Phi = \beta t. 
\eea
The remaining equations are satisfied provided, 
\be
\label{VI0_cond}
2 p_1 p_2 + p_2^2 = 4 \beta^2.  
\ee

The non-Abelian T-dual geometry follows from inverting the matrix 
\be
\gamma + \kappa = \left( \begin{array}{ccc} a_1^2 & - y  & z \\ 
 y  & a_2^2 & 0 \\
-z & 0 & a_2^2 
\end{array} \right).  
\ee
The T-dual metric and $B$-field are read off from the symmetric and anti-symmetric components of the inverse matrix, respectively, 
\bea
\dd s^2 &=&  - a_1^2 a_2^4 e^{-4 \beta t } \, \dd t^2 + \frac{1}{\Delta} \biggl( a_2^2 \dd x^2 + \left(a_1^2  + \frac{z^2}{a_2^2} \right) \dd y^2 + \frac{2 y z}{a_2^2} \dd x \dd y + \left( a_1^2  + \frac{y^2}{a_2^2} \right) \dd z^2 \biggr), \nn
B &=& \frac{1}{\Delta} \dd x \wedge (y \dd y - z \dd z),  
\eea
where we have defined 
\be
\Delta = a_1^2 a_2^2 + y^2 + z^2. 
\ee
We remark that the resulting geometry has no obvious isometries and all symmetries appear to be broken. The dilaton is again easily determined, 
\be
\Phi =  \beta t - \frac{1}{2} \ln ( a_2^2 \Delta). 
\ee
It is worth noting in this case that the trace of the structure constants is zero. For this reason, we expect a supergravity solution and it can be checked that the supergravity EOMs are satisfied, once one imposes (\ref{VI0_cond}), in line with our expectations.  

\section{Relation to Generalized Supergravity}
\label{sec:EGRSV}
In the previous section we have shown that non-Abelian T-duals of Bianchi VI$_h$ cosmological models lead to solutions to generalized supergravity where the Killing vector $I$ is the trace of the structure constants. At this stage the observation that $I$ is the trace of the structure constants may simply be a coincidence. In this section we dispel this notion by returning to the T-dual $\sigma$-model of EGRSV \cite{Elitzur:1994ri} and show that a non-local anomalous term in the $\sigma$-model captures the modification in the equations of generalized supergravity. Before doing this in general, we will present the analysis for Bianchi III. For completeness, we revisit the Bianchi V analysis of \cite{Elitzur:1994ri} in the appendix. The EOMs of generalized supergravity can be found also in the appendix. 

\subsection{Bianchi III}
In this section, we revisit the analysis of \cite{Elitzur:1994ri} but for Bianchi III. We will also work in the accustomed gauge, i. e. $g_{tt} = - 1$ and set $\Phi = 0$. Thus, we consider the Ricci-flat metric 
\bea
\label{III_metric}
\dd s^2 &=& - \dd t^2 + t^2 ( \sigma_1^2 + \sigma_3^2) + \sigma_2^2. 
\eea
where we have defined Maurer-Cartan one-forms: 
\be
\sigma_1 = \dd x, \quad \sigma_2 = \dd y, \quad \sigma_3 = e^{-x} \dd z, 
\ee
The one-forms satisfy the following differential conditions,  
\be
\dd \sigma_1 = 0, \quad \dd \sigma_2 = 0, \quad \dd \sigma_3 = - \sigma_1 \wedge \sigma_3. 
\ee
so the only structure constant is 
\be
f^{3}_{~31} = 1.
\ee

As explained previously, non-Abelian T-duality reduces to inverting the matrix, 
\be
\label{M}
\gamma + \kappa = \left( \begin{array}{ccc} t^2 & 0 & -z \\ 0 & 1 & 0 \\ z & 0 & t^2 \end{array} \right). 
\ee
The resulting solution to generalized supergravity is 
\bea
\label{III_soln}
\dd s^2 &=& - \dd t^2 + \frac{t^2}{t^4 + z^2} ( \dd x^2 + \dd z^2) + \dd y^2, \nn
B &=&  \frac{z}{t^4 + z^2} \dd x \wedge \dd z, \nn
\Phi &=& - \frac{1}{2} \ln ( t^4 + z^2).  
\eea
Here the Killing vector $I$ that completes the solution is self-selecting; although $\partial_{y}$ is Killing, we have not deformed this direction and this leaves $I = c \, \partial_{x}$, where $c$ is a constant. The correct constant of proportionality follows from the trace of the structure constant, 
\be
\label{III_vec}
I = \partial_{x}, 
\ee
and it can be checked that this constitutes a solution to generalized supergravity. We would now like to confirm that one arrives at the same constant from considering the variation of the T-dual action (\ref{dual_action2}) with respect to the background conformal factor $\sigma$, following analysis presented in \cite{Elitzur:1994ri}. 

We recall the T-dual $\sigma$-model \cite{Elitzur:1994ri}
\be
S = \frac{1}{2 \pi} \int \dd^2 z \left[ ( - \partial t \bar{\partial} t + ( \partial \tilde{X}_{j} - \tilde{c} \delta^1_{j} \partial \sigma) M^{j k} ( \bar{\partial} \tilde{X}_{k} + \tilde{c} \delta^{1}_{k} \bar{\partial} \sigma ) + \ln \det M \partial \bar{\partial} \sigma \right],  
\ee
where we have defined $\tilde{c} = f^i_{i 1}$. We can further decompose the action as follows: 
\bea
S_0 &=& \frac{1}{2 \pi} \int \dd z^2 \left( - \partial t \bar{\partial} t +  \partial \tilde{X}_{j} M^{j k } \bar{\partial} \tilde{X}_{k} +  \ln \det M \partial \bar{\partial} \sigma \right), \nn
S_1 &=& -\frac{\tilde{c}}{2 \pi}  \int \dd^2 z \sigma \left( \bar{\partial} ( M^{j 1} \partial \tilde{X}_{j} )  - \partial ( M^{1 k} \bar{\partial} \tilde{X}_{k} ) \right), \nn
S_2 &=& - \frac{\tilde{c}^2}{2 \pi} \int \dd^2 z M^{11} \partial \sigma \bar{\partial} \sigma.  
\eea
We will now consider the variation of the total action $S = S_0 + S_1 + S_2$ with respect to the conformal factor, following \cite{Elitzur:1994ri}. To leading order in $\sigma$ the variation $\delta_{\sigma} S_2 = 0$, so we will ignore this term. It is worth noting that we will not be doing a quantum calculation here, but simply importing the known one-loop result for $S_0$ and combining it with the variation of $S_1$, which is a classical contribution. It should be borne in mind that the dilaton contribution to the one-loop $\beta$-functions is also purely classical \cite{Callan:1985ia} for reasons explained in the introduction.  

The variation of $S_0$ with respect to the conformal factor gives 
\be
\pi \frac{\delta S_0}{ \delta \sigma} = \frac{1}{2} \left( \beta_{G_{\mu \nu}}^{ I=0} + \beta_{B_{\mu \nu}}^{ I=0} \right) \partial X^{\mu} \bar{\partial} X^{\nu} + \frac{1}{2} \beta_{\Phi}^{ I=0} \partial \bar{\partial} \sigma, 
\ee
where $X^{\mu} \equiv \{ t, \tilde{X}^{j} \}$ and $\beta_{B_{\mu \nu}}^{ I=0}$, $\beta_{G_{\mu \nu}}^{ I=0}$ and $\beta_{\Phi}^{ I=0}$ are the usual supergravity one-loop $\beta$-functions \cite{Callan:1985ia}, essentially the EOMs of generalized supergravity evaluated at $I=0$. From the variation of the second term $S_{1}$ with respect to $\sigma$, one finds
\bea
 \delta_{\sigma} S_1 &=&  - \frac{\tilde{c}}{2 \pi} \int \dd^2 z \delta \sigma \left[ \bar{\partial} \left( \frac{(t^2 \partial x - z \partial z)}{t^4 + z^2}\right) - \partial \left( \frac{(t^2 \bar{\partial} x + z \bar{\partial} z)}{t^4 + z^2} \right) \right] \nn
&=&  - \frac{\tilde{c}}{2 \pi} \int \dd^2 z \delta \sigma \biggl[ - \frac{2 t (t^4 - z^2)}{(t^4 + z^2)^2}  ( \bar{\partial} t \partial x - \bar{\partial} x \partial t ) + \frac{2 t^2 z}{(t^4 + z^2)^2}  ( \bar{\partial} x \partial z - \bar{\partial} z \partial x ) \nn 
&+& \frac{4 t^3 z}{(t^4 + z^2)^2} ( \bar{\partial} t \partial z + \bar{\partial} z \partial t ) + \frac{2 ( z^2 - t^2)}{(t^4 + z^2)^2}  \bar{\partial} z \partial z - \frac{2 z}{(t^4 + z^2)} \bar{\partial} \partial z \biggr].  
\eea

At this point we can use the EOM to replace the second derivative term. To leading order in $\sigma$, the EOM takes the form, 
\be
\label{EOM_S}
\partial ( M^{i j  } \bar{\partial} \tilde{X}_{j} ) + \bar{\partial} ( M^{ j i } \partial \tilde{X}_{j} ) - \frac{\delta M^{ j k } }{\delta \tilde{X}_{i}} \partial \tilde{X}_{j} \bar{\partial} \tilde{X}_{k} = 0,  
\ee
where we have focused on the spatial terms. From the EOMs, we find
\bea
\bar{\partial} \partial z &=&  \frac{2 t z}{(t^4 + z^2)} (   \bar{\partial} t \partial x - \bar{\partial} x \partial t ) \nn
&+& \frac{z}{(t^2 + z^2)} ( \bar{\partial} z \partial z - \bar{\partial} x \partial x) - \frac{(z^2 - t^4)}{t (t^4 + z^2)} ( \bar{\partial} t \partial z + \bar{\partial} z \partial t),  
\eea
and substitute it back into the above expression to get, 
\bea
\label{S1_III}
\delta_{\sigma} S_1 &=& -  \frac{\tilde{c}}{2 \pi} \int \dd^2 z \delta \sigma \biggl[  - \frac{2 t^4}{( t^4 + z^2)^2} \bar{\partial} z \partial z + \frac{2 z^2}{( t^4 + z^2)^2} \bar{\partial} x \partial x  + \frac{2 z}{t ( t^4 + z^2)} ( \bar{\partial} t \partial z + \bar{\partial} z \partial t ) \nn 
&-&  \frac{2 t }{(t^4 + z^2)} (   \bar{\partial} t \partial x - \bar{\partial} x \partial t ) + \frac{2 t^2 z }{(t^4 + z^2)^2} ( \bar{\partial} x \partial z - \bar{\partial} z \partial x) \biggr].  
\eea

Now, let us compare to the contribution due to $I = c \, \partial_{x}$ coming from the EOMs of generalized supergravity. We are particularly interested in comparing the terms that depend on the constant $c$. These contribute to the one-form $X$ (\ref{X}) in the following way: 
\be
X = \frac{c}{(t^4 + z^2)} \left( t^2 \dd x +  z \dd z \right).  
\ee
We now reproduce the EOMs of generalized supergravity focusing on the terms due to the Killing vector: 
\bea
\label{EOM_III}
\beta_{B_{t x}} &=&  -  \frac{c \, 2 t} {(t^4+z^2)}, \quad \beta_{B_{x z}} =  \frac{c \, 2 t^2 z}{(t^4 + z^2)^2}, \quad \nn
\beta_{G_{t z}} &=&   - \frac{c \, 2 z}{t (t^4 + z^2)}, \quad \beta_{G_{xx}}  =   - \frac{c \, 2 z^2}{(t^4 + z^2)^2}, \quad \beta_{G_{z z}}  = \frac{c \, 2 t^4}{(t^4 + z^2)^2}, \nn
\beta_{\Phi} &=& - \frac{4}{t^2} c ( c - 1).  
\eea
As advertised earlier, we find a solution to generalized supergravity when $c =1$, but it is interesting that the dilaton EOM is also satisfied when $I = 0$. The key point is that the terms due to the Killing vector $I$ (\ref{EOM_III}) are precisely of the same form as the terms coming from the variation of the $\sigma$-model action (\ref{S1_III}) once one sets $ c = \tilde{c}$. In other words, for this example of a non-Abelian T-dual of a Bianchi III spacetime, the variation of the $S_1$ term in the action recovers the equations of motion of generalized supergravity evaluated on the same solution provided the Killing vector $I$ is simply the trace of the structure constants. 

\subsection{General case}
It is easy to extend the analysis above to the general case. To do so, we recall that 
\be
M^{j k} = G_{j k } + B_{j k}. 
\ee
In terms of the metric and NS two-form, the equation of motion (\ref{EOM_S}) is 
\be
\bar{\partial} \partial \tilde{X}^{i} + \partial \tilde{X}^{j} \bar{\partial} \tilde{X}^{k} ( \Gamma^{i}_{j k } - \frac{1}{2} H^{i}_{~j k } ) = 0, 
\ee
where we have introduced the Christoffel symbol and field strength $H = \dd B$. Replacing $f^{j}_{ji} \rightarrow I^{i}$, we can write the anomaly term as 
\bea
\delta_{\sigma} S_1 &=& - \frac{1}{2 \pi} \int \dd^2 z \delta \sigma I^{i} \left( \bar{\partial} (M^{ji} \partial \tilde{X}^j) - \partial ( M^{ik} \bar{\partial} \tilde{X}^k )\right),  \nn
&=& - \frac{1}{2 \pi} \int \dd^2 z \delta \sigma I^{i} \biggl( 2 B_{ji} \partial \bar{\partial} \tilde{X}^j  + [ \partial_{k} G_{i j}- \partial_{j} G_{i k}  -  (\partial_{k} B_{i j} + \partial_{j} B_{i k} ) ] \partial \tilde{X}^j \bar{\partial} \tilde{X}^k \biggr), \\
&=& - \frac{1}{2 \pi} \int \dd^2 z \delta \sigma I^{i} \biggl( 2 B_{j i} \left( \frac{1}{2} H^{j}_{~k l } - \Gamma^{j}_{k l} \right) 
+  \partial_{l} G_{i k}- \partial_{k} G_{i l}  -  (\partial_{l} B_{i k} + \partial_{k} B_{i l} ) \biggr) \partial \tilde{X}^k \bar{\partial} \tilde{X}^l. \nonumber
\eea
where in the last line we have used the equation of motion. These terms precisely match the equations of motion of generalized supergravity (\ref{B_EOM}) - (\ref{Einstein}) once one uses the fact that $I$ is Killing and that equation (2.13) of \cite{Arutyunov:2015mqj}, namely the relation  
\be
I^{k} H_{k  i j } + \partial_{i} Z_j - \partial_{j} Z_i = 0, 
\ee
where we have defined $X_k = I_k + Z_k$. 

\section{Discussion}
The purpose of this note is to confirm that non-Abelian T-duality with respect to non-semisimple groups leads to solutions to generalized supergravity, in line with earlier expectations \cite{Hoare:2016wsk}. By illustrating this for Bianchi VI$_h$ spacetimes, we have extended an earlier result \cite{Fernandez-Melgarejo:2017oyu} to all non-singular Bianchi cosmology solutions to supergravity (with zero NSNS two-forms). 

Based on explicit solutions, we observed that the Killing vector is simply the trace of the structure constants. To better understand this fact, we returned to the T-dual $\sigma$-model of EGRSV \cite{Elitzur:1994ri} and noted that the variation of a non-local anomalous action with respect to the conformal factor $\sigma$ precisely matches the EOMs of generalized supergravity once the Killing vector $I$ is identified with the trace of the structure constants. While the $\sigma$-model is scale invariant, it has recently been suggested that Weyl invariance can be restored through a shift in the dilaton in a doubled formalism \cite{Sakamoto:2017wor} and it would also be interesting to understand this directly at the level of the EGRSV $\sigma$-model to see if Weyl invariance can be restored. 

Throughout this work, we have driven home the message that non-Abelian T-duality is simply a matrix inversion. We recall that all Yang-Baxter deformations can be understood as open-closed string maps \cite{Araujo:2017jkb}, which are also simple matrix inversions. It would be interesting to revisit statements in the literature connecting Yang-Baxter deformations to non-Abelian T-duality \cite{Hoare:2016wsk, Borsato:2016pas} in order to better understand this relation, especially in light of the observation that all transformations can be reduced to matrix inversions. This suggests there should exist a simple overarching description. 

Finally, it would be interesting to better understand non-Abelian T-duality. One distinctive feature of the transformation is that it decompactifies geometries, thereby obscuring the AdS/CFT interpretation. Viewing the transformation in two steps, this is easy to see why. Firstly, the metric $\gamma$ in the transformation (\ref{map}) is defined with respect to the Maurer-Cartan one-forms and not  the coordinates. This means in the case of Bianchi IX that one replaces a compact space, for example a (constant radius) three-sphere, with $\mathbb{R}^3$. The anti-symmetric terms $\kappa_{ij} = \epsilon_{ijk} x_k$, where we have used the structure constants of $SU(2)$ symmetry, simply break the translation symmetry leaving a residual $SU(2)$ symmetry from the rotation generators. In short, non-Abelian T-duality, especially for a three-sphere, looks like a deformation of a flat space geometry and not a compact geometry. It would be extremely interesting if one could decompose non-Abelian T-duality into two steps: an initial step where the original geometry is ``flattened" and a second step that determines what deformations of this flattened geometry give rise to supergravity solutions. This may shed light on various puzzling aspects of non-Abelian T-duality. 

\section*{Acknowledgements}
We acknowledge T. Araujo, N. S. Deger, D. Giataganas, M. M. Sheikh-Jabbari \& H. Yavartanoo for collaboration on related topics. We thank  B. Hoare, D. C. Thompson \& J. Nian for discussion, as well as Y. Lozano \& K. Yoshida for comments on preliminary drafts. 

\appendix 

\section{Generalized Supergravity EOMs} 
For completeness we review the EOMs of generalized supergravity \cite{Arutyunov:2015mqj}. For the purposes of this paper, it is enough to restrict our attention to the NS sector. The EOMs take the form, 
\bea
\label{B_EOM}
\beta_{B_{\mu \nu}} &=& - \frac{1}{2} \nabla^{\rho} H_{\rho \mu \nu} +X^{\rho} H_{\rho \mu \nu} + \nabla_{\mu} X_{\nu} - \nabla_{\nu} X_{\mu}, \\
\label{Einstein}
\beta_{G_{\mu \nu}} &=& R_{\mu \nu} - \frac{1}{4} H_{\mu \rho \sigma} H_{\nu}^{~ \rho \sigma} + \nabla_{\mu } X_{\nu} + \nabla_{\nu} X_{\mu} , \\
\label{dilaton_EOM} \beta_{\Phi} &=& R - \frac{1}{12} H^2 + 4 \nabla_{\mu} X^{\mu} - 4 X_{\mu} X^{\mu}, 
\eea
where we have defined 
\be
\label{X}
X_{\mu} \equiv \partial_{\mu} \Phi + (g_{\nu \mu} + B_{\nu \mu} ) I^{\nu}. 
\ee
It should be noted that usual supergravity is recovered when $I = 0$. 

\subsection{Bianchi V} 
In this section, we dissect the calculation presented in \cite{Elitzur:1994ri} to recast the terms that cancel the usual supergravity $\beta$-function as the modification inherent in generalized supergravity. Here, the T-dual solution is \cite{Elitzur:1994ri}
\bea
\dd s^2 &=& - \dd t^2 + \frac{t^2}{4 x ( t^4 + x)} \dd x^2 + \frac{x}{t^2} \dd y^2 + \frac{t^2}{t^4 + x} \dd z^2, \nn
B &=& \frac{1}{2 ( t^4 + x)} \dd x \wedge \dd z, \quad \Phi = - \frac{1}{2} \ln \left( t^2 ( t^4 + x)\right), 
\eea
where the coordinates are related to the original Lagrange multipliers as follows: 
\be
\tilde{X}_1 = z, \quad \tilde{X}_2 = \sqrt{x} \cos y, \quad \tilde{X}_3 = \sqrt{x} \sin y. 
\ee
Here the structure constants are $f_{12}^2 = f_{13}^3 = 1$, so that equating $I$ with the trace, $I^{i} = f_{ji}^{j}$, we get $I^{1} = -2$. Recalling that $\tilde{X}_1 = z$, we expect that $I = -2 \partial_{z}$ and it is easy to confirm that this is a solution to generalized supergravity. 

Let us now consider the T-dual $\sigma$-model (\ref{dual_action2}) and we again take $f^{j}_{j 1} = \tilde{c}$. To repeat the analysis, one just needs the matrix 
\be
M = \frac{1}{t^2 ( t^4 + x)} \left( \begin{array}{ccc}
t^4 & - t^2 \sqrt{x} \cos y & - t^2 \sqrt{x} \sin y  \\
t^2 \sqrt{x} \cos y & t^4 + x \sin^2 y & - x \cos y \sin y \\
t^2 \sqrt{x} \sin y & - x \cos y \sin y & t^4 + x \cos^2 y
 \end{array} \right).  
\ee
Evaluating the variation $\delta_{\sigma} S_1$, the result is \cite{Elitzur:1994ri}
\bea
\delta_{\sigma} S_1 &=& - \frac{\tilde{c}}{2 \pi} \int \dd^2 z \delta \sigma  \biggl[ - \frac{2 t}{(t^4 + x)} ( \bar{\partial} t \partial z - \bar{\partial} z  \partial t ) - \frac{t^2}{(t^4 + x)^2} (\bar{\partial} x \partial z - \bar{\partial} z  \partial x  ) \nn
&-& \frac{1}{(t^4 + x)^2} \left( \frac{1}{t} (\bar{\partial} t \partial x + \bar{\partial}  x \partial t ) - \frac{t^4}{2 x} \bar{\partial} x \partial x + 2 x \bar{\partial} z \partial z \right) + \frac{2 x}{t^4} \bar{\partial} y \partial y  \biggr], 
\eea
where once again we have used the EOM to eliminate a second derivative. Now, assuming $I = c \, \partial_{z}$, let us compare to the EOMs of generalized supergravity. To do so, we note the contribution of $I$ to $X$, 
\be
X  =  - \frac{c}{2 (t^4 + x)} \left(  \dd x - 2 t^2 \dd z \right),  
\ee
which allows us to identify the contribution to the EOMs of generalized supergravity: 
\bea
\beta_{B_{t z}} &=& -  \frac{2 c \, t}{(t^4 +x)}, \quad \beta_{B_{x z}} =  -\frac{c \, t^2}{(t^4 + x)^2}, \quad \nn
\beta_{G_{t x}} &=&   \frac{c}{t (t^4 + x)}, \quad \beta_{G_{xx}}  =   - \frac{c \, t^4}{2 x (t^4 + x)^2}, \quad \beta_{G_{z z}}  = \frac{c \, 2 x }{(t^4 + x)^2}, \quad \beta_{G_{yy}}  = - \frac{c \, 2 x}{t^4}, \nn
\beta_{\Phi} &=& - \frac{4}{t^2} c ( c + 2). 
\eea
One can quickly see that all the terms agree with $ c= \tilde{c} = -2$ and that a solution to generalized supergravity exists. Although it was noted that $\delta_{\sigma} S_1$ cancelled the contribution from the one-loop $\beta$-functions \cite{Elitzur:1994ri}, it was not appreciated at the time that the additional terms agree with the EOMs of generalized supergravity \cite{Arutyunov:2015mqj} once the Killing vector $I$ is the trace of the structure constants.

\end{document}